# Universal superconducting precursor in the cuprates

G. Yu[1,*], D.-D. Xia[1,2], D. Pelc[1,3], R.-H. He[4], N.-H. Kaneko[5], T. Sasagawa[6], Y. Li[7], X. Zhao[1,2], N. Barišić[3,8,*], A. Shekhter[9,#], and M. Greven[1,*]

[1] School of Physics and Astronomy, University of Minnesota, Minneapolis, Minnesota 55455, USA
[2] State Key Lab of Inorganic Synthesis and Preparative Chemistry, College of Chemistry, Jilin University, Changchun 130012, Zhejiang, China
[3] Department of Physics, Faculty of Science, University of Zagreb, Bijenička cesta 32, HR-10000, Zagreb, Croatia
[4] Westlake Institute for Advanced Study, Hangzhou, China
[5] National Institute of Advanced Industrial Science and Technology (AIST), 1-1-1 Umezono, Tsukuba, Ibaraki 305-8563, Japan
[6] MSL, Tokyo Institute of Technology, Kanagawa 226-8503, JAPAN
[7] International Center for Quantum Materials, School of Physics, Peking University, Beijing 100871, China
[8] Institute of Solid State Physics, TU Wien, 1040 Vienna, Austria
[9] Pulsed Field Facility, National High Magnetic Field Laboratory, Los Alamos National Laboratory, Los Alamos, NM 87545, USA
# Present address: National High Magnetic Field Laboratory, Tallahassee, FL 32310, USA
* yug@umn.edu, barisic@ifp.tuwien.ac.at, greven@umn.edu

**The nature of the superconducting (SC) precursor in the cuprates has been the subject of intense interest[1-2], with profound implications for both the normal and the SC states. Different experimental probes have led to vastly disparate conclusions on the temperature range of superconducting fluctuations[3-14]. The main challenges have been to separate the SC response from complex normal-state behavior, and to distinguish the underlying behavior of the quintessential $CuO_2$ layers from compound-specific properties. Here we reveal remarkably simple and universal behavior of the SC precursor using torque magnetometry, a unique thermodynamic probe with extremely high sensitivity to SC diamagnetism. We comprehensively study four distinct cuprate compounds: single-$CuO_2$-layer $La_{2-x}Sr_xCuO_4$ (LSCO), $Bi_2(Sr,La)_2CuO_{6+\delta}$ (Bi2201) and $HgBa_2CuO_{4+\delta}$ (Hg1201), and double-layer $Bi_2Sr_2Ca_{0.95}Y_{0.05}Cu_2O_{8+\delta}$ (Bi2212). Our approach, which focuses on the nonlinear diamagnetic response, completely removes normal-state contributions and thus allows us to trace the diamagnetic signal above $T_c$ with great precision. We find that SC diamagnetism vanishes in an unusual, yet surprisingly simple exponential manner, marked by a universal temperature scale that is independent of compound and $T_c$. We discuss the distinct possibility that this unusual behavior signifies the proliferation of SC clusters as a result of the intrinsic inhomogeneity known to be an inherent property of the cuprates.**

High-$T_c$ superconductivity in the cuprates emerges from a metallic state that exhibits unusual pseudogap phenomena[1]. One of the pivotal open questions is how superconductivity evolves from this complex metallic state. For the extensively studied systems LSCO, Bi2201 and $YBa_2Cu_3O_{6+\delta}$ (YBCO), some experiments (Nernst effect[3,4], torque magnetization[5,6], photoemission[7], infrared spectroscopy[9]) seem to indicate signatures of superconductivity in an anomalously wide temperature range above $T_c$ in the underdoped part of the phase diagram. Yet other experiments (microwave[10], terahertz conductivity[11,12], and specific heat[13]) reveal signatures of incipient superconductivity only in a relatively narrow temperature range near $T_c$. Furthermore, extensive recent work has focused on the interplay between superconductivity and other incipient ordering tendencies, including the possibility of the simultaneous appearance of superconducting and charge-density-wave (CDW) fluctuations relatively far above $T_c$ (ref. 1-2). One of the experimental problems is that it is difficult to disentangle the SC response from these other ordering tendencies, e.g, the Nernst signal can be strongly affected by CDW fluctuations[15]. Therefore, it has been challenging to reliably establish the normal-state "background" and hence to unambiguously extract the SC signal. Recent measurements of the nonlinear conductivity, which is zero in the normal state, provide further evidence that traces of superconductivity indeed vanish rapidly above $T_c$, in an exponential fashion[16]. Such temperature dependence of the nonlinear conductivity is incompatible with prevailing theoretical ideas, but consistent with an inhomogeneous SC gap distribution, i.e., with local superconductivity above $T_c$ and with SC percolation[16].

Here we use torque magnetometry to study the precursor SC diamagnetism above $T_c$ (we quote $T_c$ values that were obtained in the limit of zero magnetic field). Torque magnetometry is a thermodynamic probe with extremely high sensitivity to SC diamagnetism, a fundamental characteristic of superconductivity[17]. Upon considering the nonlinear magnetic field dependence, we find that the SC diamagnetism can be unambiguously disentangled from the normal-state paramagnetism over wide temperature ranges. By investigating a wide range of cuprate compounds over extensive doping and temperature ranges, we are thus able to carry out a pivotal test of scenarios of SC pre-pairing.

We investigate nonlinear diamagnetism in LSCO, Bi2201, Bi2212 and Hg1201, with a focus on the doping dependence of Hg1201 and on a direct comparison of all four compounds near optimal doping. Hg1201 is a model single-layer compound[18-20] that features a simple tetragonal crystal structure, minimal extrinsic disorder effects, and and an optimal transition temperature $T_c^{max}$ of nearly 100 K. We demonstrate that the emergence of superconductivity in all four cuprates is nearly indistinguishable, despite the dramatically lower $T_c^{max}$ values of LSCO and Bi2201 (just below 40 K[18]) and the double-layer nature of Bi2212. The nonlinear diamagnetism follows exponential-like rather than power-law temperature dependence, in a universal (nearly compound independent) manner, in excellent agreement with the recent complementary linear and nonlinear conductivity results[16].

In torque magnetometry, the magnetization **M** is deduced from the mechanical torque **τ** = $V\mu_0$(**M** × **H**) experienced by a crystal in an external magnetic field **H**. Here, $\mu_0$ is the permeability of free space and $V$ is the sample volume. The torque is measured as a function of temperature ($T$), magnetic field strength ($H$), and orientation of the sample with respect to the field direction. For a tetragonal system such as Hg1201 (and for nearly tetragonal systems such as LSCO, Bi2201, and Bi2212 – see Methods), the sample orientation is parameterized by the angle $\theta$ between **H** and the crystallographic $c$-axis. The field dependence of the magnetization can be obtained either directly, through field scans at a fixed angle, or indirectly, by observing the angular dependence of the torque in a fixed field. In the angular scans, the linear-in-field magnetization (paramagnetic or diamagnetic) reveals itself as the second harmonic in the angular dependence, $\tau \propto H^2\sin(2\theta)$, whereas the non-linear-in-field magnetization introduces higher harmonics. For clarity, we define the "torque susceptibility" $\chi_{torque}(H,\theta,T)$ and "torque magnetization" $M_{torque}(H,\theta,T)$ via $\chi_{torque} \equiv \tau/(V\mu_0 H_a H_c) = M_c/H_c - M_a/H_a$ ($\propto \tau/(H^2\sin(2\theta))$) and $M_{torque} \equiv \chi_{torque}H$, where $H_a = H\sin(\theta)$ and $H_c = H\cos(\theta)$ are the components of **H** along the crystallographic $a$ and $c$ directions, and $M_c$ ($M_a$) is the magnetization component along the $c$-axis (in-plane) direction. In the linear response regime, $\chi_{torque}(T)$ equals the susceptibility anisotropy $\chi_c(T) - \chi_a(T)$, and is independent of field strength and orientation of the crystal. For non-linear-in-field response, such as SC diamagnetism, $\chi_{torque}$ varies with $H$ and $\theta$ (Fig. 1c).

In the normal state, at sufficiently high temperatures, we find perfect linear-in-field paramagnetic response in both the field and angular scans up to 14 T, the highest field of our study, for all samples. This is demonstrated for Hg1201 in Fig. 1a-c, both via direct observation of the magnetization as well as via consideration of $\chi_{torque}$, and indicative of the absence of any SC diamagnetism. In contrast to the normal-state response, the SC diamagnetism manifests itself via the nonlinear magnetic field dependence, even in relatively low fields (Fig. 1c). For example, for a Hg1201 crystal with $T_c \approx 96$ K, the nonlinear magnetization in the field scans and the higher harmonics in the angular scans are clearly discernible below the same temperature of about 120 K (Figs. 1a and 1b).

The temperature dependence of $\chi_{torque}$ for Hg1201, Bi2201 and LSCO (Figs. 1d, 2a and Supplementary Figs. S3, and S4) reveals two distinct behaviors above $T_c$. The near-$T_c$ regime is clearly dominated by SC diamagnetism and features an approximately exponential decay, accompanied by a strong non-linear field dependence of the magnetization (Fig. 1a-c). At higher temperatures, $\chi_{torque}$ exhibits qualitatively different temperature dependence and no magnetic field dependence (within our sensitivity limit), and thus is clearly identified as normal-state paramagnetism. Previous torque studies of LSCO and Bi2201[5,6] deduced SC diamagnetism by subtracting an assumed high-temperature $T$-linear paramagnetic background $\chi(T) = (a + bT)$. However, we find that the normal-state paramagnetism strongly deviates from this assumption (see Fig. 1d).

Without resorting to any assumptions regarding the paramagnetic background, we identify SC diamagnetism from the difference ($\Delta_H\chi_{torque}$) of $\chi_{torque}$ at two different fields (or, equivalently, at two different angles), which completely removes the paramagnetic (i.e., linear response) component and leaves only non-linear magnetism. This enables us to trace the SC diamagnetism at temperatures well above $T_c$, even when the SC signal is two orders of magnitude smaller than the high-temperature paramagnetic magnetization (Fig. 2b-c). As shown in Figs. 2b,c, we find that the SC diamagnetic signal ($\Delta_H\chi_{torque}$) exhibits a rapid exponential decrease with increasing temperature, $\Delta_H\chi_{torque} \propto \exp\{-(T-T_c)/T_d\}$, where $T_d$ is a measure of the temperature range over which SC traces are significant. Figure 2b,d demonstrates for Hg1201 that $T_d$ exhibits weak doping dependence in the large range from the very underdoped ($p \approx 0.07$) to the overdoped ($p \approx 0.18$) part of the phase diagram.

Remarkably, our complementary measurements of optimally-doped Bi2201 and moderately underdoped LSCO reveal that the diamagnetic response of all three single-layer compounds is nearly indistinguishable (Figs. 2c-d, S2, S3), despite the stark difference (a factor of about 2.5) in $T_c^{max}$ and the prominent charge/spin "stripe" correlations in $x = 0.125$ LSCO[1]. The universal nature of the observed behavior is further demonstrated for double-layer Bi2212 ($T_c \approx 90$ K; Fig. 2c,d).

These results are significant for a number of reasons. First, they constitute an unequivocal thermodynamic determination of SC emergence in the cuprates, as we observe SC emergence directly via diamagnetism, a fundamental and prominent characteristic of superconductivity[17], and because our experimental approach does not resort to any "background" estimation. Second, they indicate that the emergence of superconductivity ($\chi_{torque}$) exhibits highly unusual, yet robust exponential temperature dependence with a characteristic temperature ($T_d$) that is clearly independent of $T_c$. Contrary to the interpretation of recent torque results for YBa$_2$Cu$_3$O$_{6+x}$[21], this behavior cannot be described by Ginzburg-Landau theory, in which $T_c$ is the characteristic fluctuation temperature scale and which predicts an approximately power-law temperature dependence[16]. Third, we demonstrate that the scale $T_d$ is universal (compound-independent). In particular, near optimal doping, the nonlinear diamagnetic response of all four investigated cuprate families is characterized by exponential decay with $T_d = 4$-$5$ K. Fourth, this is the case even for $p = 0.125$ LSCO, where stripe correlations are particularly prominent, and for Hg1201 in the $p = 0.07$-$0.11$ range, where CDW correlations are prominent in this compound[22,23]. This implies that these CDW correlations are not directly relevant to SC emergence and that they must have inadvertently affected a range of prior results[3-8]. Moreover, we note that in quantum-critical-point scenarios of the curpates, SC pairing is mediated by the fluctuations of a distinct order parameter[1]. Yet the observed behavior for overdoped Hg1201 near the putative quantum critical point at $p \approx 0.19$ is the same as at low doping. Last, but not least, the observation of an exponential decay of the SC response with a universal characteristic temperature scale is fully consistent with recent nonlinear conductivity results[16]. This implies the existence of an underlying characteristic temperature (or energy scale) proportional to $T_d$ and independent of $T_c$.

How can we understand the unusual SC emergence? We note that the cuprates are lamellar, perovskite-derived materials that are intrinsically inhomogeneous at the nanoscale[24-30], and that even simple-tetragonal Hg1201 exhibits considerable variation in local electric field gradients, Cu-O bond angles and bond distances[24,27]. Evidence for inhomogeneity is observed on multiple energy scales, ranging from about 0.1 eV in scanning tunneling microscopy (STM) to $10^{-7}$ eV in nuclear magnetic resonance measurements. STM demonstrates that both the pseudogap[28-30] and the lower-energy SC gap[29,30] exhibit considerable spatial inhomogeneity. Consequently, some of the spatially inhomogeneous SC gaps "survive" in form of SC clusters at temperatures well above $T_c$. As the temperature decreases, these clusters proliferate and grow in size, and eventually percolate near $T_c$. The emergence of superconductivity may therefore be thought of as a percolation process, with a temperature scale controlled by the distribution of the SC gap rather than by $T_c$. Quantitative evaluation of the nonlinear torque signal is difficult in this scenario, as it would involve the temperature and magnetic-field dependence of the SC gap distribution, the size distribution of the SC clusters, and the Josephson coupling among the clusters. However, we note that the recent conductivity measurements show universal exponential behavior as well, and that in this case the comparison with a simple percolation model is greatly simplified because small isolated SC clusters do not contribute to the conductance and large clusters dominate the response. The linear paraconductivity exhibits exponential decay with a characteristic temperature that is nearly identical to $T_d$ from our torque measurements, and both linear and nonlinear (third-harmonic) conductivity can be quantitatively described by a simple percolation model[16]. This demonstrates that the exponential temperature dependence can indeed be described by a percolation process. Notably, such a description fits well into a recently-proposed overarching picture of cuprate physics, where the key element is intrinsic localization-gap inhomogeneity[31]. The inhomogeneous (de)localization of charge carriers naturally accounts for the normal-state phase diagram, pseudogap, and superfluid density. The SC gap distribution discussed here can then be understood as a manifestation of the same underlying inhomogeneity on a lower, emergent energy scale.

**Methods**

**1. Material growth and characterization.** The Hg1201 crystals were grown as previously described[19]. The samples were annealed in oxygen rich or poor atmosphere to achieve the desired doping levels[20]. The optimally-doped Bi2201 (onset $T_c$ = 35 K) and underdoped Bi2212 (onset $T_c$ = 90 K) crystals were grown by the traveling-solvent floating-zone technique[18]. The quoted $T_c$ values for all samples were determined from zero-field-cooled susceptibility measurements in 5 Oe magnetic field oriented along the crystallographic $c$-axis using Quantum Design, Inc., MPMS instruments. For Bi2201 and Bi2212, $p$ is estimated[18] from the empirical relation $T_c/T_c^{max} = 1 - 82.6\,(p - 0.16)^2$. For LSCO, we use $p = x$.

**2. Torque measurements.** The torque measurements were carried out with high-sensitivity torque lever chips, using Quantum Design, Inc., PPMS instruments. Torque was detected through the resistance change of piezoresistive elements in the Wheatstone bridge on the chips. The sensitivity of the method is limited by the signal-to-noise ratio of the Wheatstone bridge. Whereas the signal is determined by the sample characteristics and sample mass, and increases with field (e.g., proportional to $H^2$ in the paramagnetic state), the noise is determined by the chip characteristics and the measurement (sampling) time. From the signal-to-noise ratio of the field dependence of the effective moment $M_{torque}$ (Fig. 1a) we estimate that our method has a sensitivity comparable to the prior torque study of LSCO and Bi2201 that used a capacitive cantilever[5,6].

In addition to the random noise in the data, the extracted nonlinear magnetic signal $\Delta_H\chi_{torque}$ is further limited by systematic error (due to imperfections in the experimental setup) from the subtraction. Therefore, the temperature dependence of $\Delta_H\chi_{torque}$ can be described by adding a small constant to the exponential decay to account for the temperature-independent systematic error. This constant results from the small errors in setting the angles, from the offset due to gravity, and from the magnetoresistance of the torque lever chip, which varies between experiments and constraint the detection limit of $\Delta_H\chi_{torque}$. As a result, the temperature range of the exponential decay appears to be cut off (Fig. 2c).

LSCO, Bi2201 and Bi2212 have nearly tetragonal distorted crystal structures. However, since the difference in planar lattice constants $a$ and $b$ is very small (for example, less than ~ 0.4% for Bi2201), the difference between $\chi_a$ and $\chi_b$ is negligible compared to that between $\chi_a$ and $\chi_c$. Therefore, these cuprates can be approximated as tetragonal systems and the same analysis for Hg1201 can be applied.

**3. Determination of $T_d$.** The characteristic decay rate $T_d$ was extracted from plots of $-1/d\ln(-\chi_{torque})/dT$ vs. $T$ (e.g., Fig. S2c). In Fig. 2d, we show $T_d$ obtained with this method: we use the values of $-1/d\ln(-\chi_{torque})/dT$ at or very close to $T_c$, where the paramagnetic contribution $\chi_p(T)$ is negligible. At low magnetic field, where Ginzburg-Landau-like behavior is expected, $\chi_{torque}$ deviates from the exponential decay. At high fields, the SC diamagnetism weakens, and the paramagnetic component $\chi_p$ becomes significant. We find that $T_d$ is determined most reliably in the moderately-high-field regime $\mu_0 H_c = 2$ to $5$ T, and we use $\mu_0 H = 3.1$-$3.4$ T in Fig. 2d. The temperature $T_d(H)$ increases with increasing field. However, this field dependence does not affect the monotonic doping dependence of $T_d$ for Hg1201 or the observed universal behavior.

**4. Contour plots.** For the contour plots in Figs. 2d, S5 and S6 we used the angular difference instead of the magnetic field difference of $\chi_{torque}$. With our experimental setup, angular scans are less prone to systematic error; all data are obtained with the same field, which minimizes the systematic error from the magnetoresistance of the chip. Furthermore, data at different angles were obtained from a single scan, whereas changing the applied field takes a relatively long time and introduces an additional error due to a small signal variation with time. This signal variation observed over long periods of time is caused by the small drift in the resistivity of the elements of the bridge, and it is effectively cancelled in the angular difference, because data for $\theta = 45°$ and $67.5°$ were taken nearly simultaneously. We also performed full angular scans at intermediate temperatures (as shown in Fig. 1b) to evaluate the higher harmonics using Fourier analysis. Since a complete angular scan is relatively time consuming, it is not practical for a detailed measurement of the temperature dependence. Measurements at two angles were sufficient for the determination of the contour plots.

We normalize $\Delta_H\chi_{torque}$ by its value at a high temperature, because data were collected with multiple chips. The normalization removes the systematic error in the absolute value of $\chi_{torque}$ due to chip calibration as well as the error in the determination of the sample mass. We use the highest temperature (250 K) measured for sample OP96 for normalization. This temperature lies far above the onset of SC diamagnetism. For all samples measured to 400 K, the monotonic temperature dependence of $\chi_{torque}$ was found to continue at higher temperatures (Fig. 1d), so that the contour plot in Figs. 2d and 3 do not depend on the particular choice of reference temperature.

The contour plots (Figs. 2d, S6) for Hg1201 are based on measurements of samples at seven doping levels with onset $T_c = 54$, $63$, $67$, $81$, $89$, $96$ and $90$ K. The $T_c = 90$ K sample is slightly overdoped. The data for the $T_c = 63$, $89$ and $90$ K samples were obtained at a lower 9 T field using a 9 T (rather than 14 T) Quantum Design, Inc., PPMS instrument. In order to combine the 9 T and 14 T data, we multiplied the 9 T results by factors of 0.79 ($\Delta_H\chi_{torque}/\chi_{torque}^{250K}$) and 1.18 ($T - T_c$). This normalization leads to a nearly perfect match of the 14 T and 9 T data for the UD67 sample.

**Acknowledgements**
We thank P.A. Crowell for the use of the 9 T Quantum Design, Inc., Physical Properties Measurement System (PPMS). The 14 T data were obtained with a PPMS in the Geballe Laboratory for Advanced Materials at Stanford University. The work on Hg1201 was funded by the Department of Energy through the University of Minnesota Center for Quantum Materials, under DE-SC-0016371 and DE-SC-0006858. The work on Bi2201 and LSCO was supported by NSF Grant No. 1006617 and by the NSF through the University of Minnesota MRSEC under Grant No. DMR-1420013. The work at the TU Wien was supported by FWF project P27980-N36 and the European Research Council (ERC Consolidator Grant No 725521).

**References**

1. Keimer, B., Kivelson, S. A., Norman, M. R., Uchida, S. & Zaanen, J. From quantum matter to high-temperature superconductivity in copper oxides. *Nature* **518,** 179-186 (2015).
2. Fradkin, E., Kivelson, S. A. & Tranquada, J. M. Colloquium: Theory of intertwined orders in high temperature superconductors. *Rev. Mod. Phys.* **87,** 457 (2015).
3. Xu, Z. A., Ong, N. P., Wang, Y., Kakeshita, T. & Uchida, S. Vortex-like excitations and the onset of superconducting phase fluctuation in underdoped $La_{2-x}Sr_xCuO_4$. *Nature* **403**, 486-488 (2000).
4. Wang, Y., Li, L. & Ong, N. P. Nernst effect in high-$T_c$ superconductors. *Phys. Rev. B* **73,** 024510 (2006).
5. Wang, Y., Li, Lu, Naughton, M. J., Gu, G. D., Uchida, S. & Ong, N. P. Field-enhanced diamagnetism in the pseudogap state of the cuprate $Bi_2Sr_2CaCu_2O_{8+\delta}$ superconductor in an intense magnetic field. *Phys. Rev. Lett.* **95,** 247002 (2005).
6. Li, L., Wang, Y., Komiya, S., Ono, S., Ando, Y., Gu, G. D. & Ong, N. P. Diamagnetism and Cooper pairing above $T_c$ in cuprates. *Phys. Rev. B* **81,** 054510 (2010).
7. Kondo, T., Hamaya, Y., Palczewski, A. D., Takeuchi, T., Wen, J. S., Xu, Z. J., Gu, G., Schmalian J. & Kaminski, A. Disentangling Cooper-pair formation above the transition temperature from the pseudogap state in the cuprates. *Nature Physics* **7,** 21-25 (2011).
8. Rourke, P. M. C., Mouzopoulou, I., Xu, X., Panagopoulos, C., Wang, Y. Vignolle, B., Proust, C., Kurganova, E. V., Zeitler, U., Tanabe, Y., Adachi, T., Koike, Y. & Hussey, N. E. Phase-fluctuating superconductivity in overdoped $La_{2-x}Sr_xCuO_4$. *Nature Physics* **7,** 455-458 (2011).
9. Dubroka, A., Rössle, M., Kim, K. W., Malik, V. K., Munzar, D., Basov, D. N., Schafgans, A. A., Moon, S. J., Lin, C. T., Haug, D., Hinkov, V., Keimer, B., Wolf, Th., Storey, J. G., Tallon, J. L. & Bernhard, C. Evidence of a precursor superconducting phase at temperatures as high as 180 K in $RBa_2Cu_3O_{7-\delta}$ (R=Y,Gd,Eu) superconducting crystals from infrared spectroscopy. *Phys. Rev. Lett.* **106,** 047006 (2011).

10. Grbić, M. S., Barišić, N., Dulčić, A., Kupčić, I., Li, Y., Zhao, X., Yu, G., Dressel, M., Greven, M. & Požek, M. Microwave measurements of the in-plane and c-axis conductivity in $HgBa_2CuO_{4+\delta}$: Discriminating between superconducting fluctuations and pseudogap effects. *Phys. Rev. B* **80,** 094511 (2009).

11. Bilbro, L. S., Aguilar, R. V., Logvenov, G., Pelleg, O., Božović, I. & Armitage, N. P. Temporal correlations of superconductivity above the transition temperature in $La_{2-x}Sr_xCuO_4$ probed by terahertz spectroscopy. *Nature Physics* **7,** 298-302 (2011).

12. Nakamura, D., Imai, Y., Maeda, A. & Tsukada, I. Superconducting fluctuation investigated by THz conductivity of $La_{2-x}Sr_xCuO_4$ thin films. *J. Phys. Soc. Jpn.* **81,** 044709 (2012).

13. Wen H.-H., Mu G., Luo, H., Yang, H., Shan, L., Ren, C., Cheng, P., Yan, J. & Fang, L. Specific-heat measurement of a residual superconducting state in the normal state of underdoped $Bi_2Sr_{2-x}La_xCuO_{6+\delta}$ cuprate superconductors. *Phys. Rev. Lett.* **103,** 067002 (2009).

14. Chang, J., Doiron-Leyraud, N., Cyr-Choinière, O., Grissonnanche, G., Laliberté, F., Hassinger, E., Reid, J-Ph., Daou, R., Pyon, S., Takayama, T., Takagi, H. & Taillefer, L. Decrease of upper critical field with underdoping in cuprate superconductors. *Nature Physics* **8,** 751–756 (2012).

15. Cyr-Choinière, O., Daou, R., Laliberté, F., LeBoeuf, D., Doiron-Leyraud, N., Chang, J., Yan, J.-Q., Cheng, J.-G., Zhou, J.-S., Goodenough, J. B., Pyon, S., Takayama, T., Takagi, H., Tanaka, Y. & Taillefer L. Enhancement of the Nernst effect by stripe order in a high-$T_c$ superconductor. *Nature* **458**, 743-745 (2009).

16. Pelc, D., Vučković, M., Grbić, M. S., Požek, M., Yu, G., Sasagawa, T., Greven, M. & Barišić, N. Emergence of superconductivity in the cuprates via a universal percolation process. Preprint.

17. Kivelson, S. A. & Fradkin, E. H. Fluctuation diamagnetism in high-temperature superconductors. *Physics* **3,** 15 (2010).

18. Eisaki, H., Kaneko, N., Feng, D. L., Damascelli, A., Mang, P. K., Shen, K. M., Shen, Z.-X. & Greven, M. Effect of chemical inhomogeneity in the bismuth-based copper oxide superconductors. *Phys. Rev. B* **69,** 064512 (2004).

19. Zhao, X., Yu, G., Cho, Y.-C., Chabot-Couture, G., Barišić, N., Bourges, P., Kaneko, N., Li, Y., Lu, L., Motoyama, E. M., Vajk, O. P. & Greven, M. Crystal growth and characterization of the model high-temperature superconductor $HgBa_2CuO_{4+\delta}$. *Adv. Mater.* **18,** 3243-3247 (2006).

20. Barišić, N., Li, Y., Zhao, X., Cho, Y.-C., Chabot-Couture, G., Yu, G. & Greven, M. Demonstrating the model nature of the high-temperature superconductor $HgBa_2CuO_{4+\delta}$. *Phys. Rev. B* **78,** 054518 (2008).

21. Kokanović, I., Hills, D. J., Sutherland, M. L., Liang, R. & Cooper, J. R. Diamagnetism of $YBa_2Cu_3O_{6+x}$ crystals above $T_c$: evidence for Gaussian fluctuations. *Phys. Rev. B* **88,** 060505(R) (2013).

22. Tabis, W., Yu, B., Bialo, I., Bluschke, M., Kolodziej, T., Kozlowski, A., Blackburn, E., Sen, K., Forgan, E. M., Zimmermann, M. v., Tang, Y., Weschke, E., Vignolle, B., Hepting, M., Gretarsson, H., Sutarto, R., He, F., Le Tacon, M., Barišić, N., Yu, G. & Greven, M. Synchrotron x-ray scattering study of charge-density-wave order in $HgBa_2CuO_{4+\delta}$. *Phys. Rev. B* **96,** 134510 (2017).

23. Hinton, J. P., Thewalt, E., Alpichshev, Z., Mahmood, F., Koralek, J. D., Chan, M. K., Veit, M. J., Dorow, C. J., Barišić, N., Kemper, A. F., Bonn, D. A., Hardy, W. N., Liang, R., Gedik, N., Greven, M., Lanzara, A. & Orenstein, J. The rate of quasiparticle recombination probes the onset of coherence in cuprate superconductors. *Sci. Rep.* **6,** 23610 (2016).

24. Agrestini, S., Saini, N. L., Bianconi, G., and Bianconi, A. The strain of $CuO_2$ lattice: the second variable for the phase diagram of cuprate perovskites. *J. Phys. A: Math. Gen.* **36,** 9133 (2003).

25. Phillips, J. C., Saxena, A. & Bishop, A. R., Pseudogaps, dopants, and strong disorder in cuprate high-temperature superconductors. *Rep. Prog. Phys.* **66,** 2111-2182 (2003).

26. Krumhansl, J. A. Fine scale mesostructures in superconducting and other materials. *Proceedings of the Conference*, Santa Fe, New Mexico, January 13-15 (1992).

27. Rybicki, D., Haase, J., Greven, M., Yu, G., Li, Y., Cho, Y. & Zhao, X. Spatial Inhomogeneities in Single-Crystal $HgBa_2CuO_{4+\delta}$ from $^{63}Cu$ NMR Spin and quadrupole shifts. *J. Supercond. Novel Magn.* **22,** 179-183 (2009).

28. Gomes, K. K., Pasupathy, A. N., Pushp, A., Ono, S., Ando Y. & Yazdani A. Visualizing pair formation on the atomic scale in the high-$T_c$ superconductor $Bi_2Sr_2CaCu_2O_{8+\delta}$. *Nature* **447,** 569-572 (2007).

29. Alldredge, J. W., Fujita, K., Eisaki, H., Uchida, S. & McElroy, K. Universal disorder in $Bi_2Sr_2CaCu_2O_{8+x}$. *Phys. Rev. B* **87,** 104520 (2013).

30. Boyer, M. C., Wise, W. D., Chatterjee, K., Yi, M., Kondo, T., Takeuchi, T., Ikuta, H. & Hudson., E. W. Imaging the two gaps of the high-temperature superconductor $Bi_2Sr_2CuO_{6+x}$. *Nature Phys.* **3,** 802-806 (2007).

31. Pelc, D., Popčević, P., Yu, G., Požek, M., Greven, M. & Barišić, N. Unusual behavior cuprates explained by heterogeneous charge localization. Preprint.

32. Li, Y., Balédent, V., Barišić, N., Cho, Y., Fauqué, B., Sidis, Y., Yu, G., Zhao, X., Bourges, P. & Greven, M. Unusual magnetic order in the pseudogap region of the superconductor $HgBa_2CuO_{4+\delta}$. *Nature* **455,** 372-375 (2008).

33. Barišić, N., Chan, M. K., Li, Y., Yu, G., Zhao, X., Dressel, M., Smontara, A. & Greven, M. Universal sheet resistance and revised phase diagram of the cuprate high-temperature superconductors. *Proc. Natl. Acad. Sci. U.S.A.* **110,** 12235–12240 (2013).

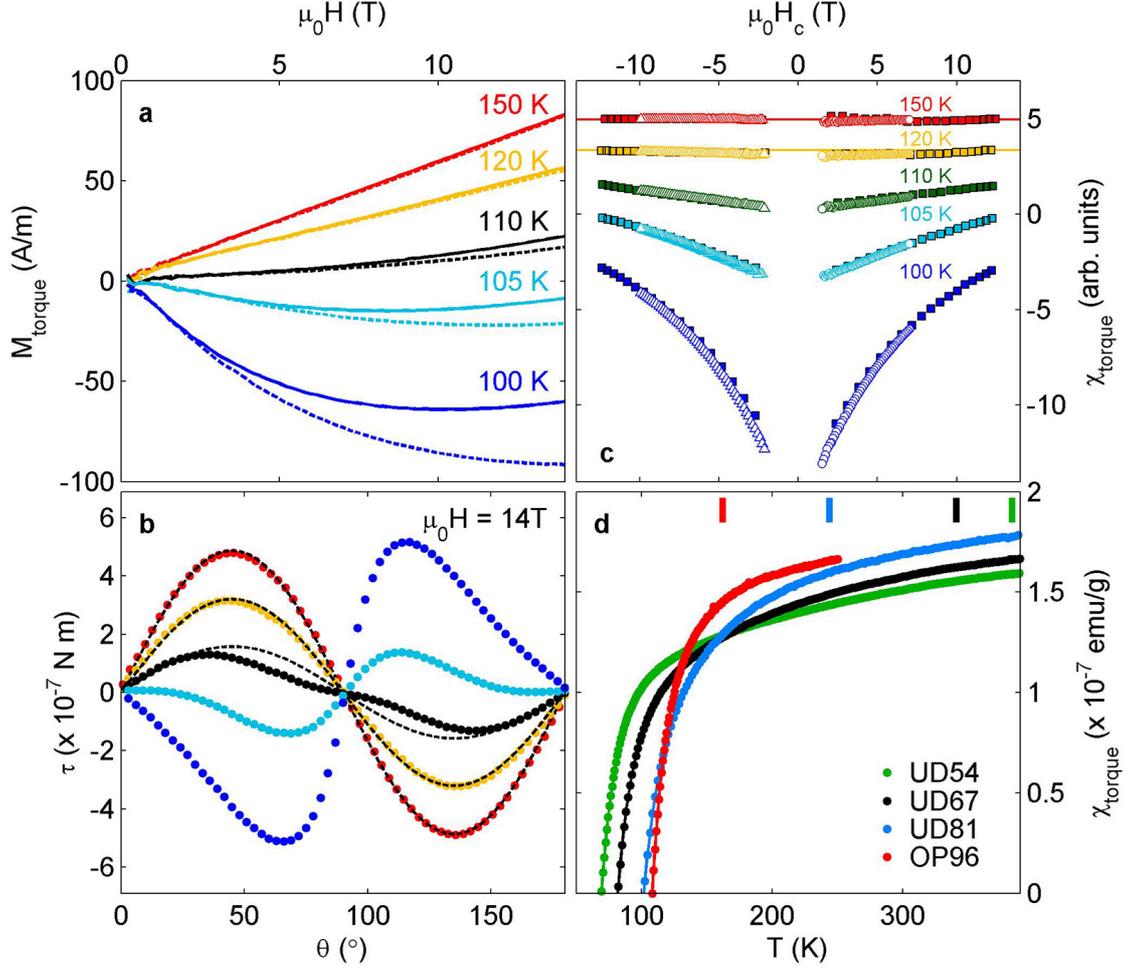

**Figure 1 | Field, temperature and doping dependence of the torque magnetization and susceptibility for Hg1201. a**, Field dependence of the effective magnetic moment $M_{torque}$ (defined in the main text) at $\theta = 45°$ (solid lines) and $60°$ (dashed lines) for a nearly optimally doped sample OP96 ($T_c \approx 96$ K). Data were taken in 0.2 T intervals. Diamagnetic magnetization is observed below about 120 K, where $M_{torque}(H)$ is strongly nonlinear. **b**, Angular dependence of the torque for sample OP96. The temperature is indicated by the same colors as in **a**. The deviation from the $\sin(2\theta)$ dependence occurs below the same temperature as the onset of the nonlinear component of $M_{torque}(H)$ in **a**. Dashed black lines are fits to $\sin(2\theta)$ for 150, 120 and 110 K. **c**, $\chi_{torque}$ as a function of $H_c = H\cos(\theta)$ for OP96, calculated from the field dependence in **a** (triangles: $\theta = 45°$; circles: $\theta = 60°$) and from the angular dependence in **b** (squares: $\mu_0 H = 14$ T). The two methods agree remarkably well, indicating that the result is hardly affected by $H_a$. This implies that the contribution from the in-plane response is negligible and that $\chi_c$ dominates the nonlinear diamagnetic signal in this temperature range. Horizontal lines at 120 K and 150 K indicate the field-independent paramagnetic contributions. **d**, $\chi_{torque}$ over a wide temperature range above $T_c$ for OP96 and three underdoped samples UD54, UD67 and UD81 ($T_c \approx 54, 67, 81$ K). $\chi_{torque}$ is obtained with an external magnetic field $\mu_0 H = 14$ T at $\theta = 45°$. At fixed field, $\chi_{torque}$ is proportional to the effective magnetization $M_{eff} \equiv \tau/(\mu_0 V H \sin(\theta))$ used in prior torque studies[5,6] (note the difference with our definition of $M_{torque}$). At high temperature, in the normal state, $M_{torque}$ is independent of $\theta$, and $\chi_{torque}$ (now equal to $\chi_c - \chi_a$) is independent of $\theta$ and $H$ up to at least 14 T

with very good accuracy. The vertical bars indicate the pseudogap temperature $T^*$ estimated from neutron scattering and planar dc resistivity measurements[32,33]. Non-trivial temperature dependences of $\chi_{torque}$ (and $d\chi_{torque}/dT$) are observed up to 400 K. Very similar behavior as in **a-d** is observed for the other cuprates (see Supplementary Information).

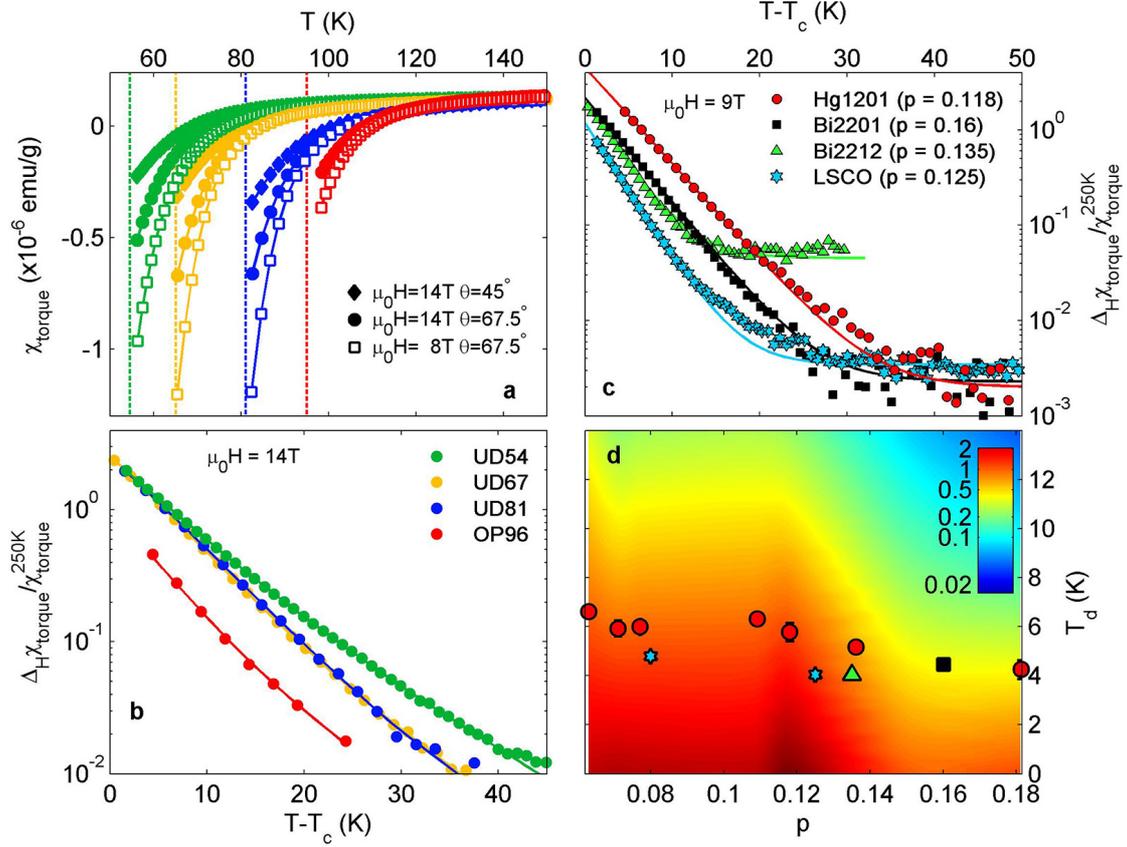

**Figure 2 | Universal behavior of the torque susceptibility. a**, $\chi_{torque}(H,\theta,T)$ obtained under three different conditions for Hg1201 samples UD54, UD67, UD81and OP96: (i) $\mu_0H = 14$ T, $\theta = 45°$; (ii) $\mu_0H = 14$ T, $\theta = 67.5°$; (iii) $\mu_0H = 8$ T, $\theta = 45°$. Deviations upon approaching $T_c$ are due to SC diamagnetism. Vertical dashed lines indicate zero-field $T_c$. $\Delta_H\chi_{torque}$ can be evaluated as either the difference between (i) and (ii), or (i) and (iii). **b**, $\Delta_H\chi_{torque}(T) \equiv \chi_{torque}(14T,45°,T) - \chi_{torque}(14T,67.5°,T)$ for Hg1201 (from **a**), normalized at 250 K, versus $T - T_c$. Lines in **a&b** are guides to the eye. **c**, Comparison of $\Delta_H\chi_{torque} \equiv \chi_{torque}(9T,45°,T) - \chi_{torque}(9T,67.5°,T)$ for slightly underdoped Hg1201 ($p = 0.118$, $T_c = 89$ K), LSCO ($p = x = 0.125$, $T_c = 27$ K), Bi2212 ($p = 0.135$, $T_c = 90$ K) and optimally doped Bi2201 ($p = 0.16$, $T_c = 35$ K). Solid lines are fits to an exponential behavior plus a small constant corresponding to the sensitivity limit (see Methods). **d**, Characteristic temperature of SC diamagnetism $T_d$ vs. hole concentration for Hg1201 (circles), Bi2201 (square), Bi2212 (triangle) and LSCO (stars) extracted for $\mu_0H_c$ in the range 3.1-3.4 T from the exponential behavior $\chi_{torque} \propto \exp\{-(T-T_c)/T_d\}$ near $T_c$. $T_d$ is nearly independent of compound and decrease slightly with increasing doping. The errors represent the uncertainty of the fits (see Methods). The color contour shows the magnitude of the nonlinear magnetic response $\log_{10}(\Delta_H\chi_{torque}/\chi^{250}_{torque})$ for Hg1201 as function of $T-T_c$ and $p$. Normalization of $\Delta_H\chi_{torque}$ by its high-temperature value allows the comparison of samples with different hole concentrations (see Methods). $T_d$ weakly increases with increasing $H_c$, but this does not affect the observation of universal behavior. Whereas $T_d$ is best defined from $\chi_{torque}$, the difference $\Delta_H\chi_{torque}$ used in Figs. 2 and Supplementary Figs. S2-S3 better demonstrates the exponential decay to high temperatures and gives a better estimate of the measurable extent of SC traces.

Supplementary Information for

"Universal superconducting precursor in the cuprates"

G. Yu, D.-D. Xia, D. Pelc, R.-H. He, N.-H. Kaneko, T. Sasagawa, Y. Li, X. Zhao, N. Barišić, A. Shekhter, and M. Greven

**Content**

Supplementary Text

1. Comparison with previous torque measurements

2. Comparison with other measurements

3. Disorder in the cuprates

Supplementary References

Supplementary Figures S1-S6

**Supplementary Text**

**1. Comparison with previous torque measurements**
In prior torque studies of LSCO and Bi2201[5,6], diamagnetism was assigned to be the deviation of the magnetization (**M** − **M**$_p$) from an assumed paramagnetic background **M**$_p$(**H**,$T$) = ($a$ + $bT$)**H**, extrapolated from the high-temperature behavior. 'Onset' temperatures of the fluctuation diamagnetism were estimated from data obtained in a 14 T field. These characteristic temperatures are substantially higher than those estimated in our study using both 9 T and 14 T fields. Sensitivity differences between the two methods are not expected to result in very different 'onset' temperatures due to the exponentially-decreasing diamagnetic signal (Figs. 2b&c, S1-5). Instead, the differences are the result of distinctly different methods of estimating the normal-state contribution. As shown for Hg1201 in Fig. 1d (and Figs. S3&S4 for Bi2201 and LSCO), the high-temperature magnetization cannot in fact be reliably approximated by a linear temperature dependence. Large deviations from assumed linear behavior already appear at or above the pseudogap temperature $T^*$, far above $T_c$ (Fig. 1d). This conclusion is also reached from considering the temperature derivative of the torque susceptibility: similar to Hg1201 (Fig. S1c&S2b), the paramagnetic susceptibility of Bi2201 (Fig. S3b) and LSCO (Fig. S4b) does not follow a linear temperature dependence, as evidenced by the fact that d$\chi_{torque}$/d$T$ is not constant in any portion of the measured temperature ranges.

Instead, we observe a clear crossover about 20 K above $T_c$ between two different temperature and field behaviors of the torque susceptibility (Fig. S3&S4). The near-$T_c$ regime exhibits strong magnetic-field dependence and a rapid exponential decay as a function of $T-T_c$. Since this behavior is directly connected to the SC transition, it is unambiguously identified as SC diamagnetism. At higher temperatures, $\chi_{torque}$ exhibits different temperature dependence and no measureable magnetic field dependence. In Figs. S3c&S4c, the deviation from an attempted $T$-linear fit at higher temperature is plotted vs. temperature. This plot clearly demonstrates the presence of a crossover, although as emphasized, it is not possible to establish the temperature and character of this crossover in this manner. However, we can now safely conclude that the magnetic response defined as the deviation from an assumed high-temperature $T$-linear behavior in the prior torque studies[5,6] contains two components, the strongly field-dependent one within about 20 K above $T_c$ identified by our method, and a field-independent (within error) one. Consequently, the temperature scale extracted in the prior work is arbitrary. Rather than being indicative of SC diamagnetism close to $T_c$, as revealed by our method, the strong temperature dependence of the magnetization at higher temperatures might result from a temperature-dependent suppression of the paramagnetic susceptibility associated with the opening of the pseudogap[32,33,S1] and/or the emergence of other electronic phases[1,2,15,S2] distinct from the SC diamagnetism close to.

At temperatures close to $T_c$, we identify two types of magnetization behavior as a function of magnetic field. As shown for Hg1201 UD67 (Fig. S1a), up to $H_c \approx 1$ T, the diamagnetic magnetization may be described reasonably well by a power-law, $|M_{torque}| \propto H_c^{\beta}$ (i.e., $|\chi_{torque}| \propto H_c^{\beta-1}$), with $\beta \approx 0.6$. A similar field dependence was observed earlier in a torque study of Bi2201 and Bi2212 (referred to as 'fragile London rigidity')[6], where $\beta$ was found to decrease from 1 to values less than 0.2 close to $T_c$. For Hg1201, just above $T_c$, we find $\beta$ close to 1/2. At higher fields, we find that the magnetization increases more slowly than $|H_c|^{0.6}$, and at even higher fields, it decreases with increasing field. All discussions of SC diamagnetism in the main text refer to this latter, moderately-high-field regime.

**2. Comparison with other measurements**
Figure S6 compares our results for the single-layer compounds Hg1201, Bi2201 and LSCO with characteristic temperatures obtained with different techniques. In evaluating this figure, it should be kept in mind that, invariably, there exist sample-specific differences (hole concentration estimates, degree of chemical disorder, etc.) and different degrees of sensitivity to the emergence of superconductivity among the experimental probes. For example, our Bi2201 sample has a $T_c$ that is 5 K higher than that of the *nominally* optimally-doped sample in the specific heat measurement[13].

As shown in Fig. S6, our result for Hg1201 agrees well with microwave conductivity work[10]: the 'onset' identified with the latter technique corresponds to $\Delta\chi_{torque}/\chi_{torque}^{250K} = 0.1\text{-}0.2$. Specific heat measurements for the low-$T_c^{max}$ compound Bi2201[13] are consistent with our torque measurements. Moreover, terahertz conductivity study of LSCO thin films shows that temporal phase correlations are observable only up to about $T_c + 20$ K[11,12]. This consistency among results obtained with charge and magnetic probes reinforces the conclusion that measurable SC signal in both low- and high-$T_c^{max}$ single-layer compounds does not extend far above $T_c$.

*The temperature scale of the robust exponential behavior identified in our work is determined for different cuprate compounds using the same criterion, hence allowing*

*us to demonstrate universality.* The temperature $T_d$ characterizes the decay rate of the SC diamagnetism rather than its apparent 'onset'. This enables unambiguous and quantitative comparison of single- and double-layer compounds with low and high $T_c^{max}$. *Our approach therefore enables us to reveal unexpected universal SC emergence.*

**3. Disorder in the cuprates**

The cuprates are intrinsically disordered and exhibit considerable variation in Cu-O bond angles and bond distances[18,24-27,S3]. Even stoichiometric $YBa_2Cu_3O_7$ (YBCO) is not as 'clean' a system as often assumed: NMR measurements reveal a considerable distribution of local electric field gradients, and EXAFS data indicate a relatively large local strain of the $CuO_2$ planes[24]. The crystal structures of the cuprates consist of an intergrowth of Cu-O layers of fixed oxygen composition and oxide layers with (typically) variable oxygen concentration. In general, there exists a bond-length mismatch across the interlayer "interface," which corresponds to a deviation of the geometric tolerance factor from unity and typically leads to lower than tetragonal global structural symmetry[S4-S5]. The bond-length mismatch is further accommodated by local deviations from the average crystal structure. Additional sources of disorder include substitutional doping (as in LSCO), off-stoichiometry in the intervening layers (in most compounds), and structural domain boundaries in compounds with lower than tetragonal symmetry. As a result, there may exist rare regions with a local electronic environment that resembles the more ideal situation in which the pairing energy is high. For example, a STM study of double-layer Bi2212 revealed a (static) spatial distribution of local gaps well above $T_c$[28]. Therefore, diamagnetic islands form locally at high temperature, and then grow in area as $T_c$ is approached from above.

**Supplementary References**

S1. Varma, C. M. Theory of the pseudogap state of the cuprates. *Phys. Rev. B* **73**, 155113 (2006).

S2. Tranquada, J. M., Sternlieb, B. J., Axe, J. D., Nakamura, Y. & Uchida, S. Evidence for stripe correlations of spins and holes in copper oxide superconductors. *Nature* **375**, 561-563 (1995).

S3. Bobroff, J., Alloul, H., Ouazi, S., Mendels, P., Mahajan, A., Blanchard, N., Collin, G., Guillen, V. & Marucco, J.-F. Absence of static phase separation in the high $T_c$ Cuprate $YBa_2Cu_3O_{6+y}$. *Phys. Rev. Lett.* **89**, 157002 (2002).

S4. Goodenough, J. B. & Manthiram, A. Crystal chemistry and superconductivity of the copper oxides. *Journal of Solid State Chemistry* **88**, 115-139 (1990).

S5. Goodenough, J. B. Chemical and structural relationships in high-$T_c$ materials, *Supercond. Sci. Technol.* **3**, 26-37 (1990).

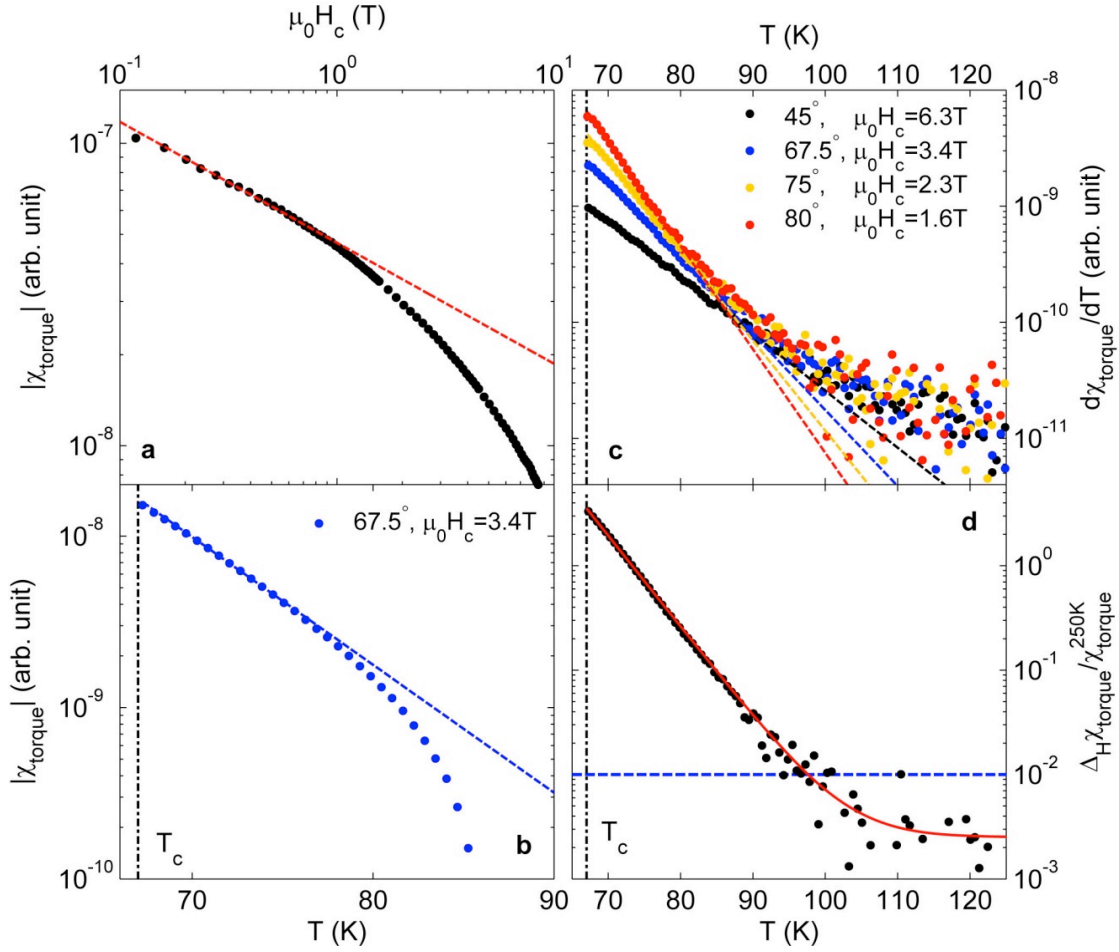

**Figure S1 | Low-field regime, definition of characteristic temperature, crossover to high-temperature regime for Hg1201 ($T_c$ = 67 K). a**, Indication of possible power-law field dependence below about 1 T for sample UD67 at 70 K, just above $T_c$. Dashed line represents a fit to $|\chi_{torque}| = |H_c|^{-0.4}$. **b**, Demonstration of exponential temperature dependence $|\chi_{torque}| \propto \exp\{-(T-T_c)/T_d\}$ at a somewhat larger field. Blue dashed line is a fit to the low-temperature data. Vertical dashed-dotted line indicates $T_c$. **c**, The exponential decay is also seen from the derivative $d\chi_{torque}/dT$, which is shown together with fits (lines) at several angles (i.e., different $H_c$). $T_d$ exhibits weak field dependence. Above about $T_c + 15$ K, a crossover to a different behavior is seen. **d**, Difference $\Delta_H\chi_{torque}(T) \equiv \chi_{torque}(45°,T) - \chi_{torque}(67.5°,T)$, normalized by the value at 250 K. The solid red line is a fit to exponential behavior, $\exp\{-(T-T_c)/T_d\}$ (with $T_d \approx 5$ K), plus a small constant to capture the noise floor. The blue horizontal dashed line indicates the limit $\Delta\chi_{torque}/\chi^{250}_{torque} = 0.01$, below which the SC diamagnetic signal can no longer be reliably discerned. The data in **b-d** were obtained with a 9 T field.

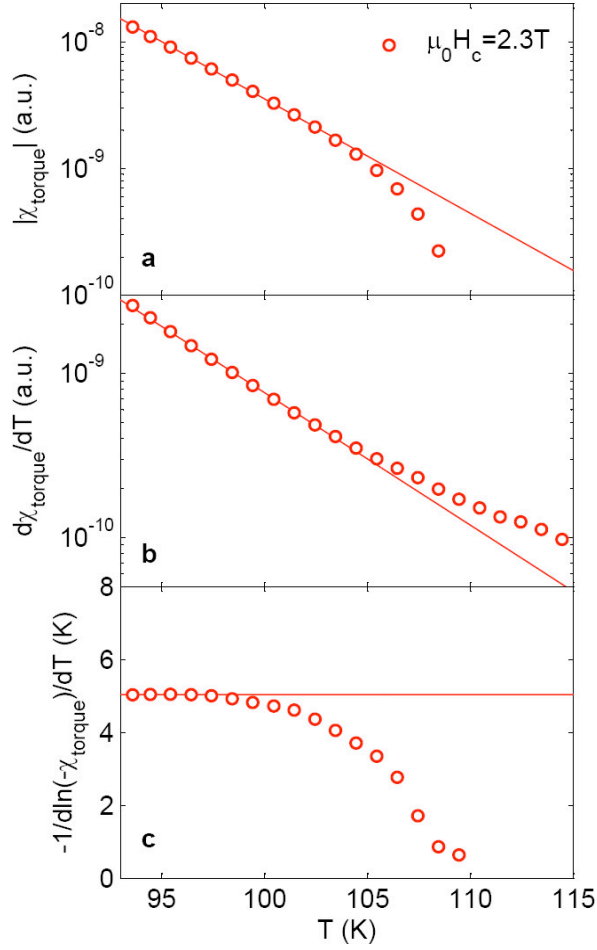

**Figure S2 | Extraction of characteristic temperature $T_d$ from $\chi_{torque}$ for Hg1201 ($T_c$ = 89 K). a**, Near $T_c$, for magnetic fields outside the Ginzburg-Landau-like regime ($H_c > 1$ T), the magnitude of $\chi_{torque}$ is well described by a simple exponential decay, $\chi_{torque} \propto \exp\{-(T-T_c)/T_d\}$ (solid line), allowing the definition of the characteristic fluctuation temperature $T_d$. The deviation from the exponential behavior above ~105 K is due to the increasing relative importance of the paramagnetic normal-state contribution. **b**, The temperature derivative $d\chi_{torque}/dT$ demonstrates the same exponential behavior as in **a** and is less sensitive to the paramagnetic contribution. **c**, Equivalently, the logarithmic derivative $-1/d\ln(-\chi_{torque})/dT$, which corresponds to the ratio of the quantities in panels **a** and **b**, equals $T_d$ near $T_c$. The deviation from the constant value ($T_d$) is due to the paramagnetic contribution that dominates at higher temperature. As shown in Fig. 3d, the doping (and compound) dependence of $T_d$ is small. We note that $T_d$ weakly increases with increasing $H_c$, but that this does not affect the fact that $T_d$ gradually decreases with increasing doping. While $T_d$ is best defined from $\chi_{torque}$, the difference $\Delta_H\chi_{torque}$ used in Figs. 2-3 and Figs. S3-S4 demonstrates the exponential decay to higher temperature and gives a better estimate of the measurable extent of superconducting diamagnetism.

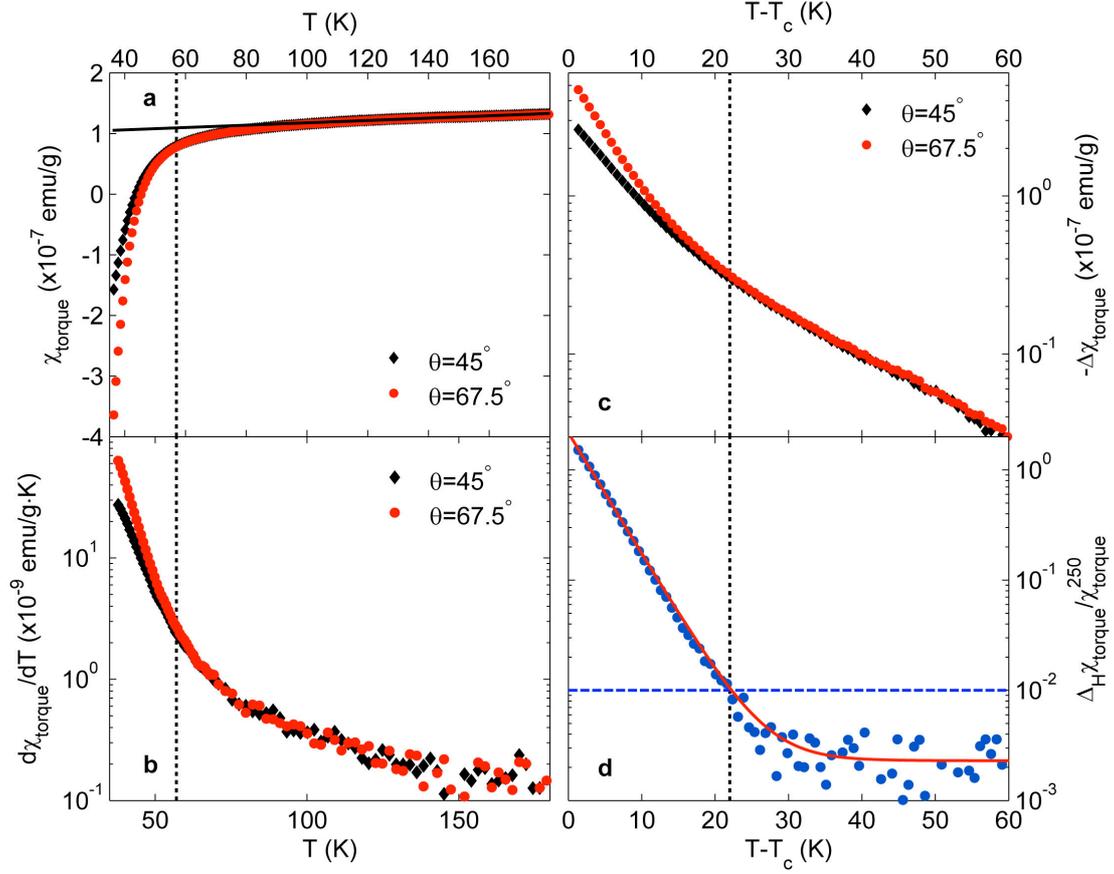

**Figure S3 | Torque susceptibility for optimally-doped Bi2201 ($T_c \approx 35$ K).** Data were taken with a 9 T field. **a**, $\chi_{torque}$ obtained at $\theta = 45°$ and $67.5°$, i.e., at different values of $H_c$, the magnetic field component perpendicular to the CuO$_2$ planes. The black solid line is a linear fit to the data between 100 K and 180 K. The deviation from an apparent high-temperature $T$-linear behavior was previously interpreted as due to SC diagmagnetism[5,6]. However, a field-dependent diamagnetic response is noticeable only at the lower temperature indicated by the vertical dashed line. **b**, The temperature derivative d$\chi_{torque}$/d$T$ changes continuously at all temperatures. **c**, $\chi_{torque}$ with the $T$-linear fit (defined in **a**) subtracted. The data clearly show two distinct temperature regimes. The near-$T_c$ regime is characterized by a strong magnetic field dependence and a rapid exponential decay of the diamagnetic magnetic response with increasing temperature. The higher temperature regime exhibits no field dependence within our sensitivity limit. **d**, $\Delta_H\chi_{torque}/\chi^{250}_{torque}$ from the difference between data taken at $\theta = 45°$ and $67.5°$. As in Fig. S1d, the red solid line is a fit to $\exp\{-(T-T_c)/T_d\}$ plus a small constant to capture the noise floor. The blue horizontal dashed line indicates the limit $\Delta_H\chi_{torque}/\chi^{250}_{torque} = 0.01$ below which the SC diamagnetic signal can no longer be reliably discerned. The vertical dashed line indicates the corresponding characteristic temperature.

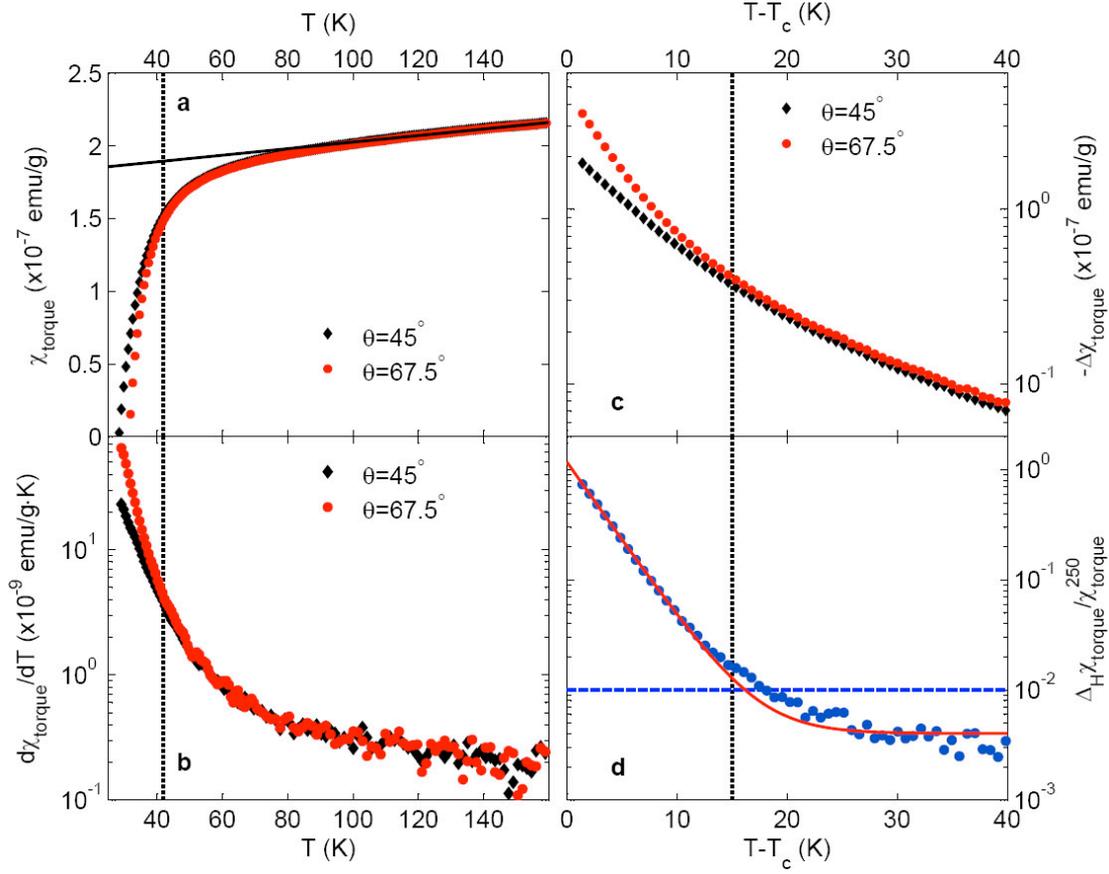

**Figure S4 | Torque susceptibility for $La_{1.875}Sr_{0.125}CuO_4$ ($T_c \approx 27$ K).** Data were taken with a 9 T field. **a**, $\chi_{torque}$ obtained at $\theta = 45°$ and $67.5°$, i.e., at different values of $H_c$, the magnetic field component perpendicular to the $CuO_2$ planes. The black solid line is a linear fit to the data between 100 K and 180 K. The deviation from an apparent high-temperature $T$-linear behavior was previously interpreted as due to fluctuation diagmagnetism[5,6]. However, a field-dependence of the magnetic response is noticeable only at much lower temperature. **b**, Temperature derivative $d\chi_{torque}/dT$ changes continuously (all the way to 350 K, the highest temperature of the measurement), indicating that there is no true $T$-linear range at this doping level. **c**, $\chi_{torque}$ with the $T$-linear fit (as defined in **a**) subtracted. The data show two distinct temperature regimes. The near-$T_c$ regime is characterized by a strong magnetic field dependence and a rapid exponential decay of the magnetic response with increasing temperature. We ascribe this regime to superconducting pre-pairing. The higher-temperature regime exhibits no field dependence within our sensitivity limit, and is thus unrelated to superconductivity. **d**, $\Delta_H\chi_{torque}/\chi^{250}_{torque}$ from the difference between data taken at $\theta = 45°$ and $67.5°$. As in Figs. S1d and S3d, the red solid line is a fit to $\exp\{-(T-T_c)/T_d\}$ plus a small constant to capture the noise floor. The blue horizontal dashed line indicates the limit $\Delta_H\chi_{torque}/\chi^{250}_{torque} = 0.01$ below which the SC diamagnetic signal can no longer be reliably discerned. The vertical dashed line in all panels indicates the corresponding characteristic temperature.

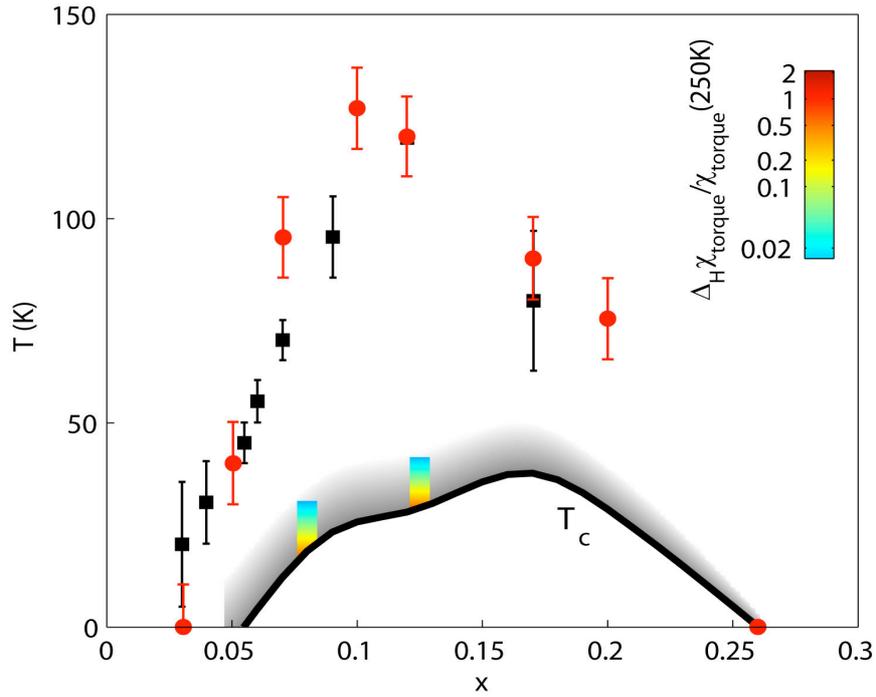

**Figure S5 | Comparison with prior torque and Nernst work for LSCO.** Pioneering Nernst effect measurements[3,4] (circles) yielded very high characteristic temperatures and were interpreted as indicative of SC fluctuations. Subsequent torque data[6] (squares) were argued to be consistent with the Nernst results. In contrast, our torque data imply a dramatically smaller regime of SC diamagnetism: the color contour shows $\log_{10}(\Delta_H \chi_{torque}/\chi^{250}_{torque})$ for the $x = 0.08$ and $0.125$ samples (see also Fig. 4). The exponentially decreasing diamagnetic signal is below our detection limit for temperatures above about $T_c + 20$ K.

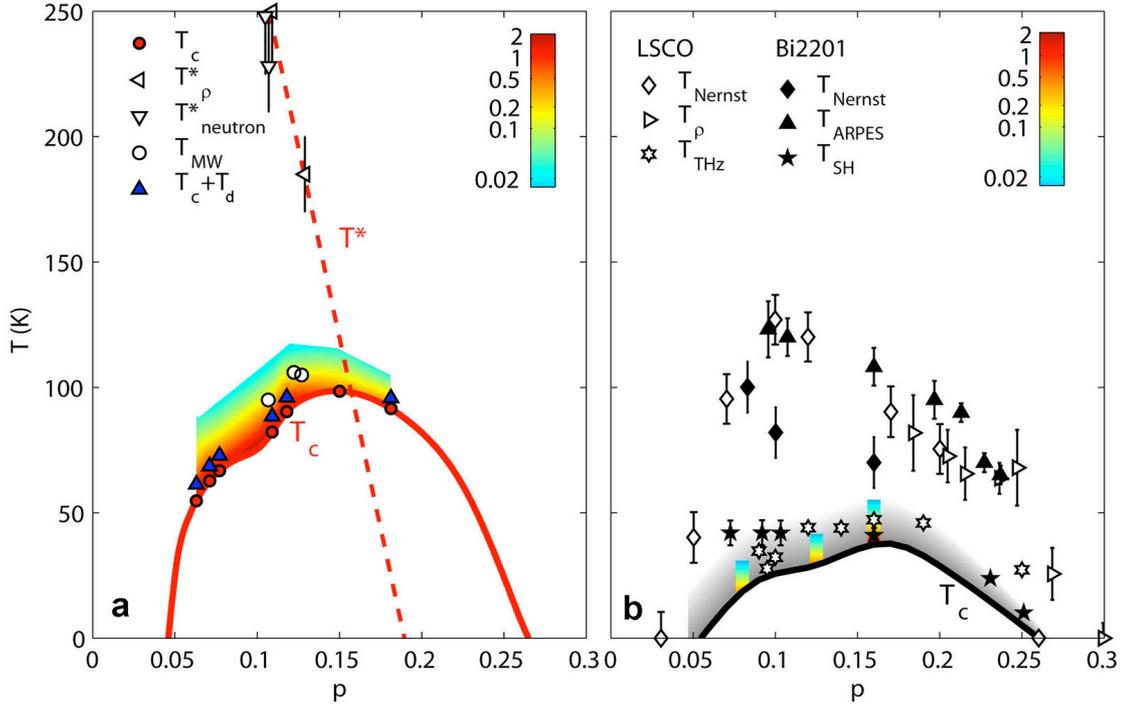

**Figure S6 | Comparison of temperature scales from various experiments.** The color contour shows $\log_{10}(\Delta_H\chi_{torque}/\chi^{250}_{torque})$ for Hg1201 (same data as in Fig. 2d, but for wider temperature range), obtained from an interpolation of measurements of seven samples, and for optimally-doped Bi2201 and underdoped LSCO ($p = x = 0.08$ and 0.125). The grey shaded area indicates schematically the extent of SC diamagnetism in LSCO and Bi2201. Similar to Hg1201 (torque and microwave[10] ($T_{MW}$)), the lower characteristic temperatures for LSCO (torque and teraherz[11] ($T_{THz}$)) and Bi2201 (torque and specific heat[13] ($T_{SH}$)) indicate that SC responses are restricted to a narrow temperature range above $T_c$, and closely track $T_c$ with doping (shaded grey area). For LSCO, the high characteristic temperatures are from Nernst ($T_{Nernst}$)[3,4] measurements, whereas for Bi2201 they are from Nernst ($T_{Nernst}$)[4] and photoemission ($T_{ARPES}$)[7] measurements. For LSCO, the characteristic temperatures from dc magnetoresistance ($T_\rho$)[8] are also shown. See Fig. S5 for a direct comparison between current and prior[5,6] torque results for LSCO. $T^*(p)$ deduced from neutron ($T^*_{neutron}$)[32] and transport ($T^*_\rho$)[20,33] measurements for Hg1201 extrapolates to zero at $p \approx 0.19$. The estimation of the hole concentrations for Hg1201 and Bi2201 is described in the Methods. For LSCO, $p = x$ is the Sr concentration. The torque data for Hg1201 (for LSCO and Bi2201) were obtained with $H = 14$ T (with $H = 9$ T).

9